\newcommand{\half}{\ensuremath{\frac{1}{2}}}

\renewcommand{\vec}[1]{\mbox{\boldmath{$#1$}}}
\newcommand{\tens}[1]{\mbox{\boldmath{$\mathsf{#1}$}}}
\newcommand{\mstag}[1][]{\ifthenelse{\equal{#1}{}}
                            {\ensuremath{\vec{M}^\dagger}}
                            {\ensuremath{\vec{M}_{#1}^\dagger}}}
\newcommand{\mferro}[1][]{\ifthenelse{\equal{#1}{}}
                            {\ensuremath{\vec{M}^F}}
                            {\ensuremath{\vec{M}_{#1}^F}}}

\newcommand{\ham}[1][]{\ensuremath{\mathcal{\hat{H}}}}

\newcommand{\nuc}[2]{\ensuremath{\mathrm{{}^{#2} #1}}} 

\newcommand{\et}{{\it et al}}

\documentclass[twocolumn,amsmath,amssymb,showpacs,prb]{revtex4}

\usepackage{graphicx}
\usepackage{dcolumn}
\usepackage{bm}

\begin{document}


\title{NMR and NQR study of the tetrahedral frustrated quantum spin system Cu$_2$Te$_2$O$_5$Br$_2$ in its paramagnetic phase}

\author{Arnaud Comment}
\email{arnaud.comment@epfl.ch}
\altaffiliation[Present address: ]{Laboratory for functional and metabolic imaging, Ecole Polytechnique F\'{e}d\'{e}rale de Lausanne, CH-1015 Lausanne, Switzerland.}
\author{Hadrien Mayaffre}
\author{Vesna Mitrovi\'{c}}
\altaffiliation[Present address: ]{Department of Physics, Brown
University, Providence, RI 02912.}
\author{Mladen Horvati\'{c}}
\author{Claude Berthier}
\affiliation{Laboratoire National des Champs Magn\'{e}tiques
Intenses, CNRS UPR3228 , Universit\'{e} J. Fourier, BP166 38042 Grenoble, France \\
 }

\author{B\'{e}atrice Grenier}
\affiliation{INAC, SPSMS, CEA, F-38042 Grenoble, France}

\author{Patrice Millet}
\affiliation{Centre d'Elaboration de Mat\'{e}riaux et d'Etudes
Structurales, CEMES/CNRS, F-31062 Toulouse, France}

\date{\today}

\begin{abstract}
The quantum antiferromagnet Cu$_2$Te$_2$O$_5$Br$_2$ was
investigated by NMR and NQR. The $\nuc{Te}{125}$ NMR investigation
showed that there is a magnetic transition around 10.5\,K at 9\,T,
in agreement with previous studies. From the divergence of the
spin-lattice relaxation rate, we ruled out the possibility that
the transition could be governed by a one-dimensional divergence
of the spin-spin correlation function. The observed anisotropy of
the $\nuc{Te}{125}$ shift was shown to be due to a spin
polarization of the 5$s^2$ ``E'' doublet of the [TeO$_3$E]
tetrahedra, highlighting the importance of tellurium in the
exchange paths. In the paramagnetic state, Br NQR and NMR
measurements led to the determination of the Br hyperfine coupling
and the electric field gradient tensor, and to the the spin
polarization of Br $p$ orbitals. The results demonstrate the
crucial role of bromine in the interaction paths between Cu spins.
\end{abstract}

\maketitle

\section{\label{sec:level1}Introduction}

In quantum antiferromagnets, triangular or tetrahedral
coordination generates strong frustration. Unusual singlet ground
states deriving from this frustration have been theoretically
predicted and actively searched for in the recent years, mainly on
Kagome or pyrochlore systems, in which the frustrated units
(triangle or tetrahedra) are sharing corners. A different type of
geometry, in which the tetrahedra units are isolated, and only
weakly coupled has been discovered with the  compounds
Cu$_2$Te$_2$O$_5$X$_2$ (X = Br, Cl) \cite{Johnsson01}, which
 contain tetrahedral clusters
of Cu$^{2+}$ ($S=\frac{1}{2}$) in a distorted square planar
CuO$_3$X coordination. These tetrahedra align to form chains along
the [001] direction, and are separated along the [100] and [010]
directions by different Te-O coordinations \cite{Lemmens01}.
Although the ground state of individual tetrahedron is expected to
be a singlet (quasi-0D system), it turns out that, below about
12\,K (18\,K for the Cl compound), the intertetrahedra couplings
lead to an incommensurate magnetic ground state with anomalous
thermodynamics properties \cite{Zaharko01,Zaharko02}. To determine
relevant dimensionality of the system several different models
were considered. These include: quasi-1D ones, assuming an
interaction between tetrahedra along the $c$-axis
\cite{Brenig01,Totsuka01,Gros01}; quasi-2D consisting of
interacting frustrated plaquettes in the $ab$-plane, in which
intertetrahedra couplings are assumed to be important
\cite{Kotov01,Whangbo01,Kotov02};  and models of a 3D tetrahedral
cluster-spin system \cite{Brenig02,Valenti01}.
 Despite all these studies,
the exact dimensionality of the system remains unclear.
Nevertheless, the results of Jensen \et. and Jagli\u{c}i\'{c} \et.
appear to favor a 3D over 1D nature of the magnetic
transition\cite{Jensen01,Jaglicic01,Jensen02}. It  is likely that
both the intratetrahedral  (and thus the frustration), leading to
a creation of spin-gaps, and the intertetrahedral interactions,
inducing a magnetic long range order, are present and compete
together.

One of the important
unsettled questions is the relative strength of
the various exchange couplings within and between tetrahedra,
which determine the dimensionality of the system. In this paper,
we present NMR and NQR measurements performed on single crystals
of Cu$_2$Te$_2$O$_5$Br$_2$. The purpose of this study was to address
 the question of the magnetic phase dimensionality via an investigation
 in the vicinity of the phase transition and to determine the spin
 polarization of the Cu ligands. One challenge that came along
 was the rather intricate assignment of the various NMR transitions
 possible in this system.

\section{\label{sec:Te NMR}Tellurium NMR}

\subsection{\label{sec:SpecStruct}Spectrum structure}

\begin{figure}
\includegraphics [width=3.4in] {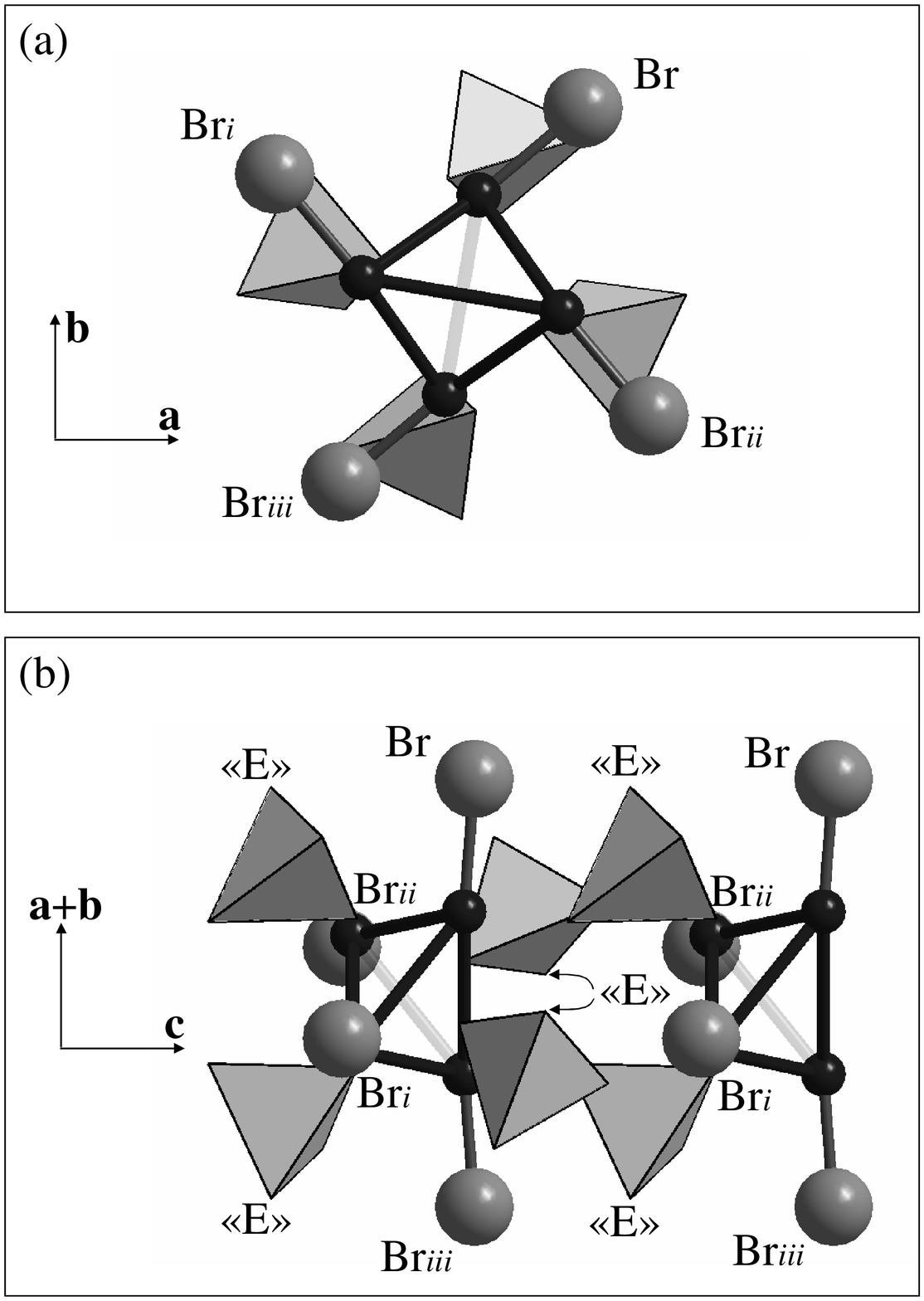}
\caption{\label{fig:Structure} Schematic views of the Cu$_2$Te$_2$O$_5$Br$_2$ structure
along the [001] axis (a) and the [$\overline{1}$10] axis (b).
 Copper atoms (small spheres) are interconnected to emphasize the tetrahedral configuration.
 Tellurium atoms (not represented) are placed inside the sketched ``O$_3$E" tetrahedra, E representing the
5s$^2$ lone pair of the Te atom\cite{Johnsson01}.}
\end{figure}

Tellurium has two NMR-active isotopes and both have a spin-$\half$
nucleus, but the natural abundance of $\nuc{Te}{125}$ is about 8
times higher than the one of $\nuc{Te}{123}$. For this reason, all
Te NMR measurements were performed on $\nuc{Te}{125}$.
Cu$_2$Te$_2$O$_5$Br$_2$ crystalizes in the P$\overline{4}$
($a=b=7.8~\AA$, $c=6.4~\AA$) space group, meaning that the
elementary pattern CuTeO$_{2.5}$Br is present four times in each
unit cell. As a consequence, although $\nuc{Te}{125}$ has a
spin-$\half$ nucleus and therefore yields to a single resonance
line, the crystal has four inequivalent Te sites for an arbitrary
orientation with respect to the direction of the external magnetic
field $B_0$. Thus, the $\nuc{Te}{125}$ NMR spectrum in
Cu$_2$Te$_2$O$_5$Br$_2$ is generally composed of four lines.
However, by applying the field in the $ab$-plane, two sites become
equivalent, while applying it along the $c$-axis, all four sites
become equivalent (see Fig.\,\ref{fig:Structure}; for a full
description of the crystal symmetry, see e.g. the article of
Johnsson \et. \cite{Johnsson01}).

\subsection{\label{sec:Hyperfine}Hyperfine shift}

\begin{figure}
\includegraphics [width=3.4in] {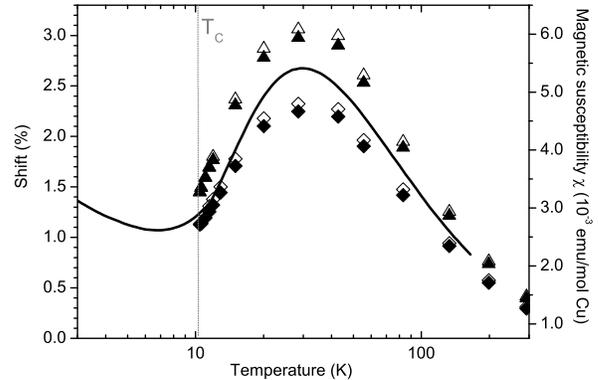}
\caption{\label{fig:ChiTe} Temperature dependence of the shift of
the four inequivalent $\nuc{Te}{125}$ nuclear spins measured in a
field of 9\,T almost parallel to [110] superimposed to the SQUID
susceptibility measurements (line). The existence of four NMR
lines is due to a slight misorientation of the field out of the
$ab$-plane}
\end{figure}

The temperature dependence of the electron spin susceptibility of
Cu$_2$Te$_2$O$_5$Br$_2$ has been extensively studied by means of
DC and AC
susceptometry\cite{Johnsson01,Lemmens01,Jensen01,Jaglicic01,Prester01}.
NMR measurements provide a way to probe the local electron spin
susceptibility through hyperfine interactions with the advantage
of being essentially insensitive to paramagnetic impurities. This
is of particular interest for probing magnetic systems at low
temperature when the contribution from paramagnetic impurities
becomes larger than the system intrinsic susceptibility. The
temperature dependence of the resonance frequencies, proportional
to the macroscopic spin susceptibility, of the four inequivalent
$\nuc{Te}{125}$ nuclear spins in the crystal  and their
temperature dependence measured in a field of 9\,T parallel to a
direction nearly parallel to [110] is shown in
Fig.\,\ref{fig:ChiTe}. The data is superimposed to the SQUID
susceptibility measurements performed with a field of 0.1 T
oriented along [110] on the same single crystal.

\begin{figure}
\includegraphics [width=3.4in] {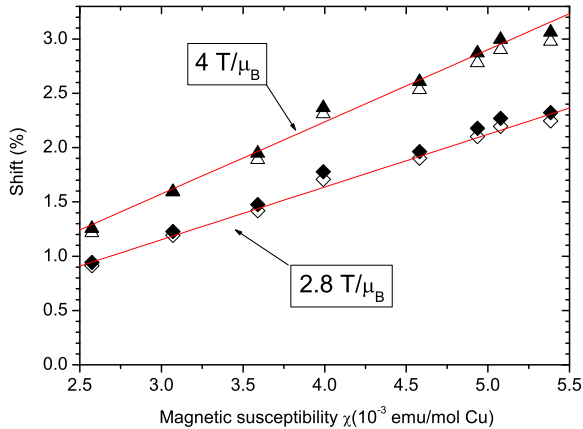}
\caption{\label{fig:KdeChiTe} Linear relationship between the
shift of the four inequivalent $\nuc{Te}{125}$ nuclear spins
measured in a field of 9\,T almost parallel to [110] and the SQUID
susceptibility. The calculation of the slopes lead to the
determination of the hyperfine field.}
\end{figure}

\begin{figure}
\includegraphics [width=3.4in] {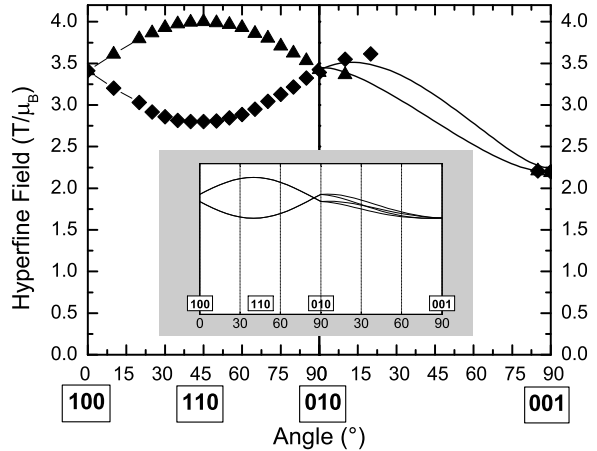}
\caption{\label{fig:AngleShift} Crystal orientation dependence of
the $\nuc{Te}{125}$ hyperfine field measured at 15\,K and 9.4\,T.
The crystal was aligned  to obtain only two resonance peaks in the
$ab$-plane (with this orientation, the filled diamonds and the
filled triangles in Fig.\,\ref{fig:ChiTe} and
Fig.\,\ref{fig:KdeChiTe} would be indistinguishable from the open
diamonds and the open triangles, respectively). Inset: dipolar
field calculated for a Te orbital pointing towards the center of
the Br-Br axis.}
\end{figure}

The data plotted in Fig.\,\ref{fig:ChiTe} allow for the
determination of the $\nuc{Te}{125}$ hyperfine coupling by
comparing the temperature dependence of the NMR frequencies to the
temperature dependence of the magnetic susceptibility as shown in
Fig.\,\ref{fig:KdeChiTe}. In doing so, we took advantage of the
multiple sites, and thus the multiple resonances, to determine the
zero-shift frequency as the extrapolated frequency at which all
sites have the same resonance frequency, in the present case
$f_0\cong$121.5\,MHz. This value corresponds to the
$\nuc{Te}{125}$ frequency for which the contribution of the Cu
electron spins polarization is zero. Note however that it is not
the resonance frequency of the ``bare" $\nuc{Te}{125}$ nuclear
spin (121.07\,MHz\cite{Orion01}) since the Te electron shells also
shift (essentially isotropically) the resonance. We observe here
that this shift is about 0.35\%, which is in the range of the
observed shifts in transition-metal tellurides\cite{Orion01}.
Conjunctively, the dependence of the NMR frequencies on the
crystal orientation (see Fig.\,\ref{fig:AngleShift}) yields to the
full determination of the hyperfine tensor. From the data shown in
Fig.\,\ref{fig:AngleShift}, we deduced that the tellurium
hyperfine shift (reflecting the spin susceptibility) is mainly
isotropic with a small anisotropic part in the $ab$-plane along a
principal axis nearly parallel to the [110] direction. Assuming an
environment of axial symmetry (i.e. neglecting a small anisotropy
in the plane perpendicular to [110]), we can define the hyperfine
shift along the external magnetic field $\vec{B}_0$ as
$K(\theta)=K_{iso}+K_{ax}(3\cos^2\theta-1)/2$, where $\theta$ is
the angle between $\vec{B}_0$ and the principal anisotropy axis of
the Knight shift tensor $\tens{K}$, $K_{iso}$ the isotropic part
of $\tens{K}$ and $K_{ax}$ its anisotropic part along the
principal anisotropy axis. The data lead to
$K_{iso}$=3.2\,T/$\mu_B$ and $K_{ax}$=0.8\,T/$\mu_B$. A simple
computation of the dipolar contribution of a single Cu electron
spin cannot account for either the amplitude, which is 8 times
weaker than the observed value, nor the angular dependence shown
in Fig.\,\ref{fig:AngleShift}. Considering transferred
polarization on Br atoms cannot account for our observations as
well. However, the measured angular dependence can be well
described by considering the contribution of a Te orbital pointing
towards the center of the Br-Br axis (see simulation shown in the
inset of Fig.\,\ref{fig:AngleShift}). Johnsson \et.  pointed out
that the Te atom is placed at the center of the O$_3$E
tetrahedron, where E represents the 5s$^2$ lone pair of the Te
atom\cite{Johnsson01}. As shown in Fig.\ref{fig:Structure}(b), the
``E'' apex of the tetrahedron stands in between two bromine atoms
along the $c$-axis and the ``E'' orbital should therefore point in
the [110] direction. Our results thus suggest that part of the
spin polarization is located in this orbital. This observation is
fully compatible with the description of Johnsson \et. who suggest
that the ``E'' orbital participates in the electronic structure
binding the two neighboring Br atoms along the
$c$-axis\cite{Johnsson01}.

\begin{figure}
\includegraphics [width=3.4in] {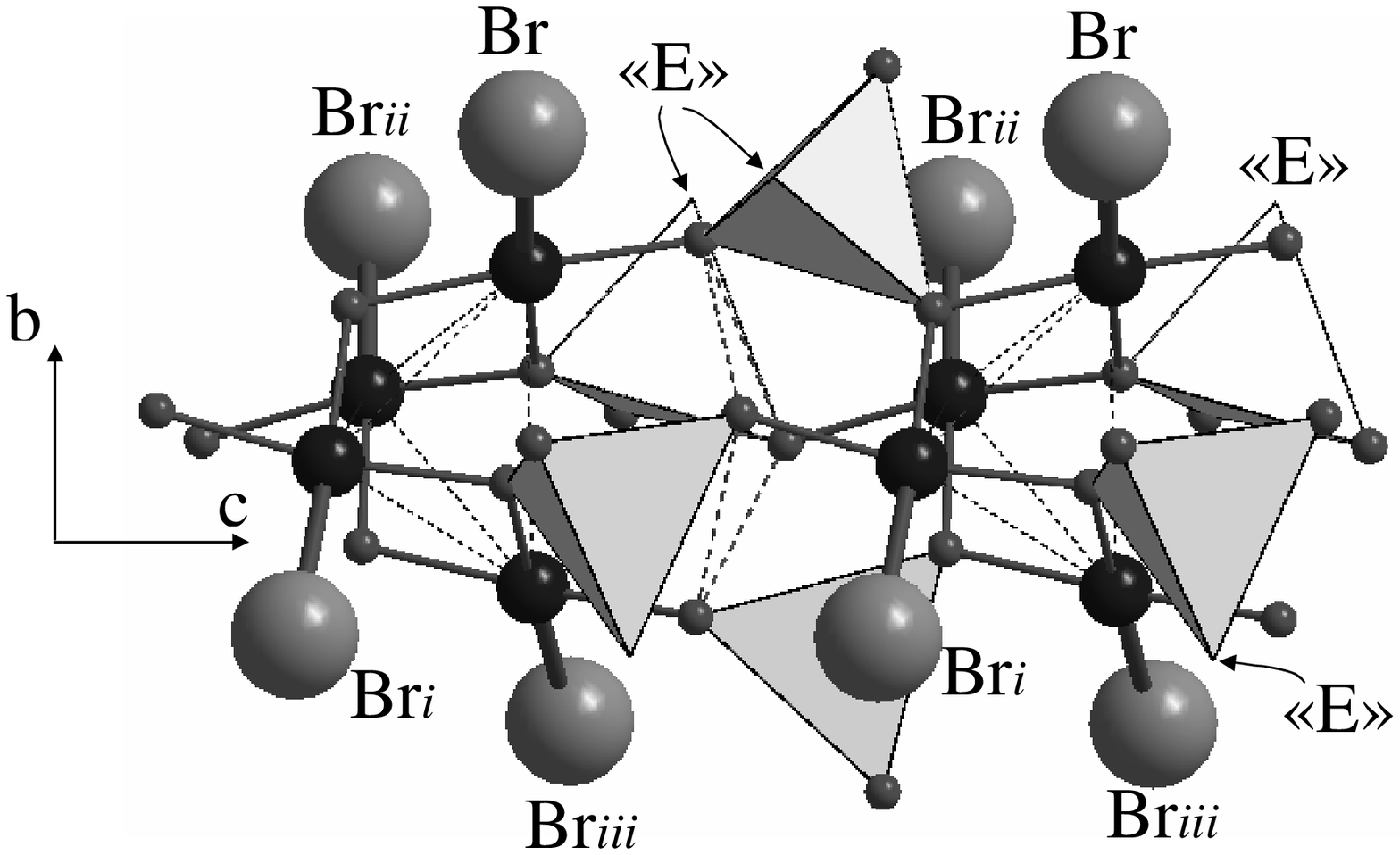}
\caption{\label{fig:StructureOx} Schematic view of the
Cu$_2$Te$_2$O$_5$Br$_2$ structure along the [100] axis.
 Oxygen atoms (smallest spheres) are displayed and dashed lines show the exchange paths proposed by \cite{Whangbo01}.}
\end{figure}

A quantitative estimation of this contribution is more delicate.
First, it should be noted that a nearly axial symmetry around the
[110] direction is not compatible with a transfer of polarization
from the three oxygens forming the tetrahedral environment of the
tellurium atom, the "E" orbital being the fourth corner. In order
to respect the symmetry, all 3 oxygen atoms should equally
contribute, which is highly unlikely as their local environment
differs dramatically from one another (see Fig.
\ref{fig:StructureOx}) . Assuming that the ``E'' orbital can be
described by a superposition of $5s$ and $5p$ orbitals (the
tetrahedral symmetry of tellurium site suggests a $sp^3$
hybridization) with one $sp^3$ orbital pointing in the [110]
direction, we can write that $K_{ax}=6/5f_p\mu_B<r^{-3}>$, where
$f_p$ is the fraction of unpaired electron in the corresponding
orbital and $<r^{-3}>$ is the mean value of $1/r^3$ over the $5p$
orbital. By taking $<r^{-3}>$=$104\cdot 10^{24}\;{\rm
cm}^{-3}$\cite{Morton01}, one finds that $f_p$=0.7\%. Similarly,
from $K_{iso}=8\pi/3f_s\mu_B<|\Psi(0)|^2>$, with $<|\Psi(0)|^2>$
being the square of the $s$-wave function at the nucleus averaged
over those electrons at the Fermi surface, one can estimate the
spin density in the $5s$ contribution to $sp^3$ orbital to give a
contact term consistent with the isotropic spin part. Knowing that
in an ideal $sp^3$ orbital the $s$ contribution is 4 times smaller
than in pure $s$ orbital, by taking the value given by
Morton\cite{Morton01}, i.e.
$<|\Psi(0)|^2>$=$\frac{1}{4}\cdot170\cdot 10^{24}$cm$^{-3}$, one
finds $f_s$=0.96\%. This value is nearly identical to the value of
$f_p$ determined from dipolar contribution, which confirms this
description in terms of $sp^3$ orbital.

In conclusion, this contribution from the ``E'' orbital well
describes the measurements. In addition,  only a small spin
polarization is needed in the Te ``E'' doublet to   quantitatively
account for the data. One should note  that this interpretation is
not compatible with the model proposed by Whangbo \et. in which
the interactions between tetrahedral clusters are presumably  from
two types of super-superexchange paths\cite{Whangbo01}: one is
Cu-O-O-Cu path in the $c$ direction  and the other Cu-Br-Br-Cu
path  in the $ab$-plane with a path Cu-Br-Br-Cu\cite{Whangbo01}.
Our results suggest that the relevant path is Cu-Br-``E''-Br-Cu in
the $c$ direction.

\subsection{\label{sec:MagneticTransition}Magnetic phase transition}

It has been reported in an earlier study that the system undergoes
a magnetic transition at a temperature $T_C$ of about 12\,K in an
external magnetic field of 9\,T\cite{Lemmens01}. In the present
study, we observe that the $\nuc{Te}{125}$ resonance line suddenly
disappears, as the temperature is lowered towards $T_C$. Although
it was possible to observe the resonance at temperatures very
close to $T_C=$12\,K, we were not able to observe the signal at
temperatures below the magnetic transition temperature. This might
be due to a significant broadening of the line, a strong
shortening of the spin-spin relaxation time, a very large
frequency shift or possibly a combination of these effects.

\begin{figure}
\includegraphics [width=3.4in] {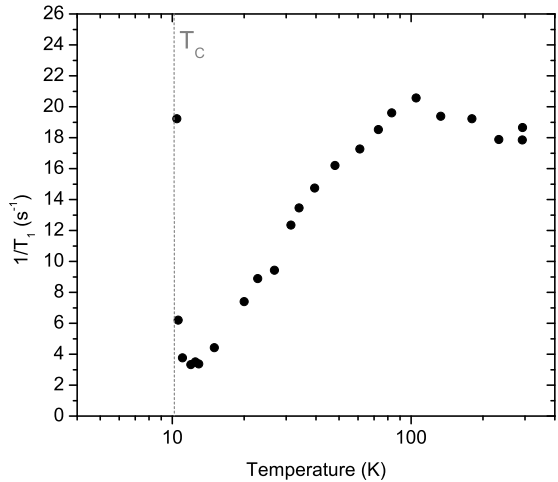}
\caption{\label{fig:R1vsT} Temperature dependence of the $\nuc{Te}{125}$
spin-lattice relaxation rate measured at 9\,T on the
lowest-frequency resonance (filled diamonds in
Fig.\,\ref{fig:ChiTe} and Fig.\,\ref{fig:KdeChiTe}).}
\end{figure}

The temperature dependence of the $\nuc{Te}{125}$ spin-lattice
relaxation time measured on the lowest-frequency resonance in a
field of 9\,T along a direction nearly parallel to [110] is shown
in Fig.\,\ref{fig:R1vsT}. The $\nuc{Te}{125}$ spin-lattice
relaxation rate decreases with decreasing temperature for
temperatures below 100\,K, but abruptly increases around 12\,K.
Such a dramatic change in spin-lattice relaxation rate is an
evidence for strong local field fluctuations and therefore for a
magnetic transition. The occurrence of a divergence in the
$\nuc{Te}{125}$ spin-lattice relaxation rate within a narrow
region of a few K above $T_C$ unambiguously shows the
three-dimensional character of the magnetic system. In a
quasi-one-dimensional system, the fluctuations would extend on a
temperature range comparable to $T_C$. The transition temperature
deduced from these measurements is $T_C=$10.5\,K, which is
slightly lower than the value determined by Lemmens
\et.\cite{Lemmens01}.

\section{\label{sec:Br NMR}Bromine NMR and NQR in the paramagnetic state}

Halogen nuclei have a large quadrupole moment and they have been
extensively studied by nuclear quadrupole resonance
(NQR)\cite{Lucken01}. NQR frequencies strongly depend on the ionic
character of the M-X bond where M is a metal ion and X is the
halogen ion\cite{Morgen01}. Both bromine isotopes have a spin-3/2
nucleus and $\nuc{Br}{79}$ and $\nuc{Br}{81}$ have almost
equivalent natural abundance (Br nuclear properties are summarized
in Table\,\ref{tab:bromine isotopes}). Cu$_2$Te$_2$O$_5$Br$_2$
contains tetrahedral arrangements of Cu atoms each one of them
placed at the center of a distorted square CuO$_3$Br. As shown in
Fig.\,\ref{fig:Structure} (see also \cite{Johnsson01}),
copper-bromine bonds are almost perpendicular to the $c$-axis of
the crystal ($90\pm 4.42$\,degrees). While Cu-Br and
Cu$_{iii}$-Br$_{iii}$ are nearly parallel to the [110] direction,
Cu$_i$-Br$_i$ and Cu$_{ii}$-Br$_{ii}$ are nearly parallel to the
[1$\overline{1}$0] one. As for tellurium, there are four
inequivalent Br sites, which reduce to two inequivalent sites when
the direction of $B_0$ is in the $ab$-plane, and to one single
site if $B_0$ is parallel to the $c$-axis.

\begin{table}
\caption{\label{tab:bromine isotopes} Spin, natural abundance,
gyromagnetic ratio,  electric quadrupole moment\cite{CRC01}, and
measured nuclear quadrupole frequencies of the two bromine
isotopes.}
\begin{ruledtabular}
\begin{tabular}{cccccc}
{} & Spin & Nat. abund. (\%) & $\gamma_n/2\pi$ (MHz/T) & $Q$ (barn) & $\nu_{NQR,15K}$ (MHz)\\
\hline
$\nuc{Br}{79}$ & 3/2 & 50.69 & 10.7 & 0.313 & 87.41\\
$\nuc{Br}{81}$ & 3/2 & 49.31 & 11.53 & 0.262 & 73.02\\
\end{tabular}
\end{ruledtabular}
\end{table}

The total Hamiltonian of a Br nuclear spin in the
Cu$_2$Te$_2$O$_5$Br$_2$ paramagnetic phase can be written as

\begin{equation}
\label{Hamiltonian}
\ham=-\gamma_n\hbar\vec{B_0}\cdot\vec{I}-\gamma_n\hbar\vec{B_0}\cdot\tens{K}\cdot\vec{I}+\frac{eQ}{2I(2I-1)}\vec{I}\cdot\tens{V}\cdot\vec{I},
\end{equation}

where the first term is the nuclear Zeeman Hamiltonian, the second
term is the hyperfine Hamiltonian with $\tens{K}$ the Knight shift
tensor, and the third term is the quadrupole Hamiltonian, in which
$e$ is the elementary charge, $Q$ is the quadrupole moment and
$\tens{V}$ is the electric field gradient (EFG) tensor. In its
principal axis coordinate system ($X,Y,Z$), the electric field
gradient is diagonal and traceless. In this particular frame,
Eq.\,\ref{Hamiltonian} can be rewritten as

\begin{equation}
\label{Simplified Hamiltonian}
\ham=-\gamma_n\hbar\vec{B_0}\cdot\vec{I}-\gamma_n\hbar\vec{B_0}\cdot\tens{K}\cdot\vec{I}+\frac{1}{6}\nu_Q[3I_Z^2-I(I+1)+\frac{\eta}{2}(I_+^2+I_-^2)],
\end{equation}

where $\eta=(V_{XX}-V_{YY})/V_{ZZ}$ is the asymmetry parameter of
the electric field gradients, and $I_+=I_X+iI_Y$ and $I_-=I_X-iI_Y$ are the spin
raising and lowering operators. For $I=3/2$,
$\nu_Q=\nu_{NQR}(1+\eta^2/3)^{-1/2}$, where $\nu_{NQR}$ is the
pure quadrupole resonance frequency.

\begin{figure}
\includegraphics [width=3.4in] {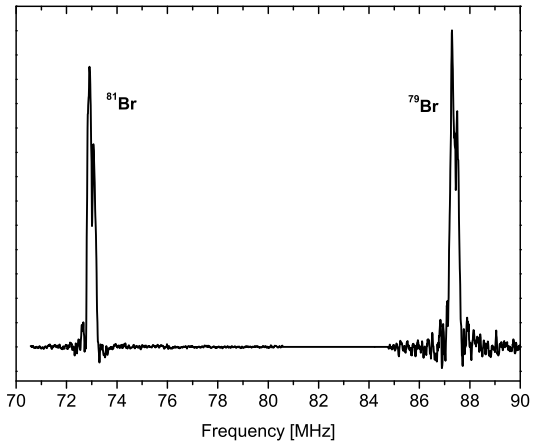}
\caption{\label{fig:NQRspectrum} Br NQR spectrum measured at
15\,K. The intensities have been divided by the square of the
frequency. Note that the slight splitting observed on both lines
is due to the presence of a residual non-zero $\vec{B_0}$ field
 in the superconducting
coil.}
\end{figure}

\begin{figure}
\begin{center}
$\begin{array}{c@{\hspace{0.2in}}c}
\includegraphics [width=1.6in] {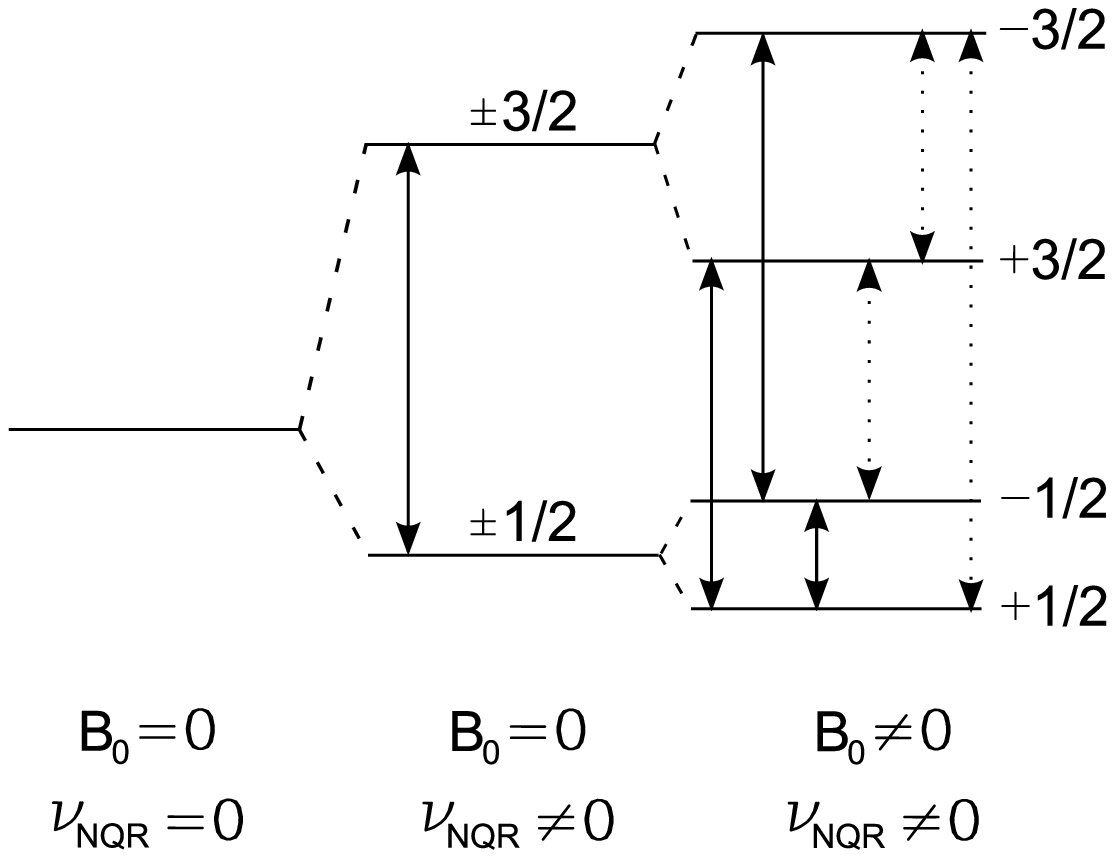} &
\includegraphics [width=1.6in] {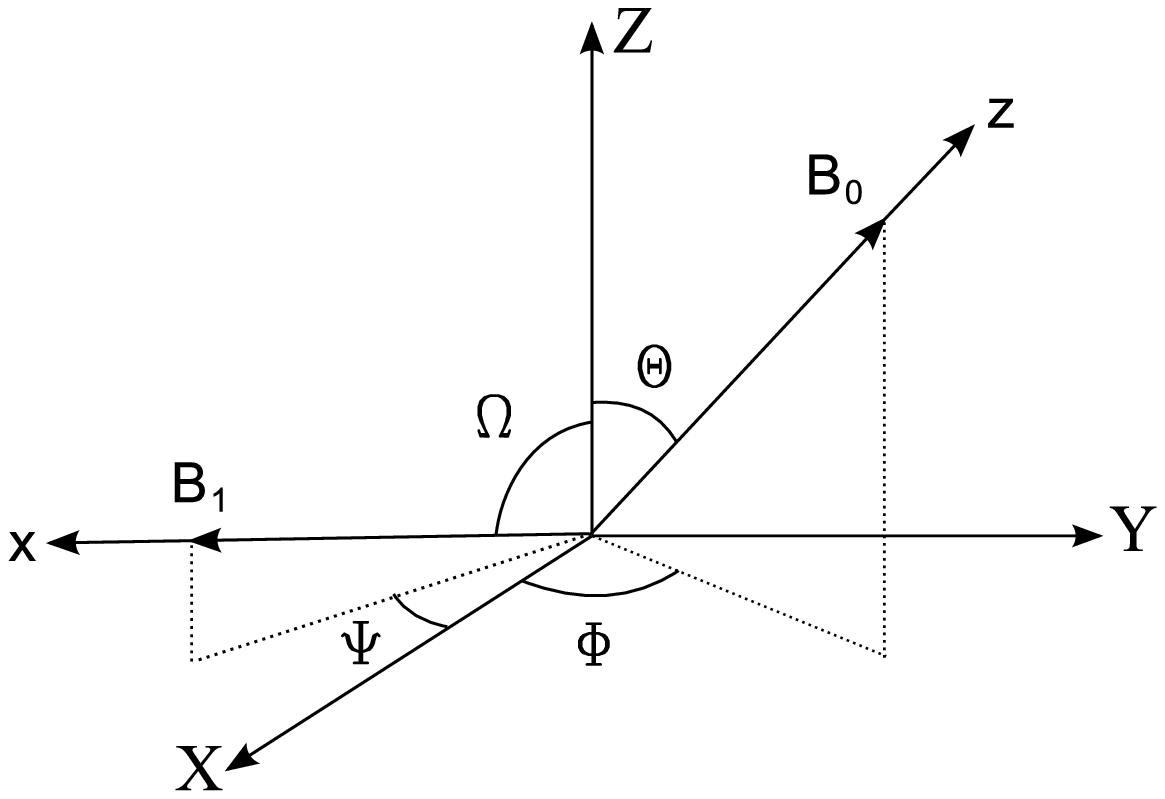} \\ [0.4cm]
\mbox{\bf (a)} & \mbox{\bf (b)}
\end{array}$
\end{center}
\caption{\label{fig:levels} (a)  Sketch of the energy levels of Br
nuclei. The solid arrows are the $\Delta m=1$ transitions and the
dotted ones the $\Delta m>1$ transitions (b) Definition of angles
$\Theta$, $\Phi$, $\Omega$, and $\Psi$. The $X$,$Y$, and $Z$ axes
correspond to the principal axes of the EFG and Knight shift
tensors. }
\end{figure}

At 15\,K and in the absence of applied static magnetic field
$\vec{B_0}$, we observed two lines of identical intensity, one at
$^{79}\nu_{NQR}$=87.41\,MHz corresponding to the $\nuc{Br}{79}$
quadrupole resonance, and the other at $^{81}\nu_{NQR}$=73.02\,MHz
corresponding to that of $\nuc{Br}{81}$ (see
Fig.\,\ref{fig:NQRspectrum}). These values are in agreement with
the ratio of the nuclear quadrupole moments published in the
literature\cite{Alonso01}. Given the gyromagnetic ratio of the two
bromine isotopes (c.f. Table\,\ref{tab:bromine isotopes}), it is
clear that the high-field approximation, which consists in
considering the quadrupole interaction as a perturbation to the
Zeeman one, will not yield to the correct transition energies for
standard NMR field values. Unlike in the case of a Zeeman only or
a quadrupole only Hamiltonian, the eigenvectors of the total
Hamiltonian are not pure and therefore the so-called forbidden
transitions can have a non-zero probability of occurring. For
$I=3/2$, the six possible transitions between the different
available spin states are shown in Fig.\,\ref{fig:levels}(a). As a
consequence, analytical solutions cannot be calculated and a
numerical computation is required.

For the present study, we developed a MATLAB routine to calculate
the field dependence of the resonance frequencies and their
associated intensities for Br sites in an arbitrary orientation of
the field. The code was written such as to numerically diagonalize
the Hamiltonian described in Eq.\,\ref{Simplified Hamiltonian},
compute the resonance frequencies from its eigenvalues and
determine the expected relative intensity of each transition by
calculating
$|<\varphi_i|\gamma_n\hbar\vec{B_1}\cdot\vec{I}|\varphi_j>|^2$,
$i\neq j$, where $\vec{B_1}$ is the radio-frequency excitation
field created in the NMR coil and $\varphi_i$, $\varphi_j$ are
eigenstates of the Hamiltonian. In the $x,y,z$ laboratory frame,
$\vec{B_1}\cdot\vec{I}$ can be expressed in terms of the $X,Y,Z$
projections of $\vec{I}$ using two Euler angles, $\Omega$ and
$\Psi$ defined in Fig.\,\ref{fig:levels}(b)), giving
$\vec{B_1}\cdot\vec{I}=B_1(I_X\sin\Omega\cos\Psi+I_Y\sin\Omega\sin\Psi+I_Z\cos\Omega)$.
The magnitude of the transition probabilities will thus strongly
depend on the intensity of $\vec{B_0}$ as well as on its direction
in the $X,Y,Z$ frame, i.e. on the crystal orientation. This is
particularly important in the present study where the Zeeman and
quadrupolar terms are of comparable magnitude.


\begin{figure}
\includegraphics [width=3.4in] {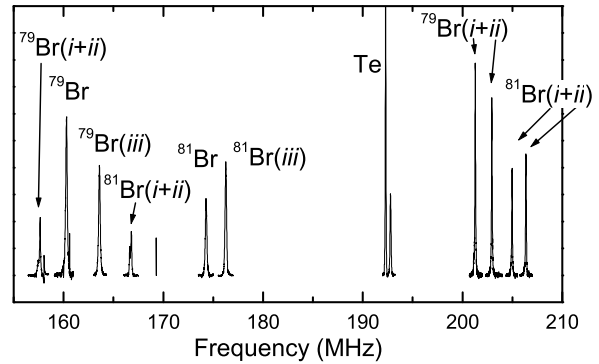}
\caption{\label{fig:BrSpectrum} NMR Spectra for $B_0=14$\,T along
[110]. The intensities have been divided by the square of the
frequency. Each line has been identified. The two sharp lines, one
around 158\,MHz and the other around 169\,MHz indicate the
position of the $\nuc{Cu}{63}$ and $\nuc{Cu}{65}$ resonance of the
copper NMR coil and have been used to determine the exact value of
$B_0$. At low frequencies (150-180\,MHz) we can find the 8 central
lines. For Br$_i$ and Br$_{ii}$ these lines are overlapping since
shift and quadrupolar frequency are small in this orientation of
the field and the misorientation is not sufficient to separate
them. At high frequencies (200-210\,MHz), we observe only the four
high frequency satellites of Br$_i$ and Br$_{ii}$. Low frequency
satellites are expected below 100\,MHz and Br and Br$_{iii}$ high
frequency satellites are expected around 240\,MHz. Note that since
the field direction is close to the principal axis of the EFG
tensor, the contribution of the quadrupolar term to the resonance
frequency is nearly maximum). }
\end{figure}

\begin{figure}
\includegraphics [width=3.4in] {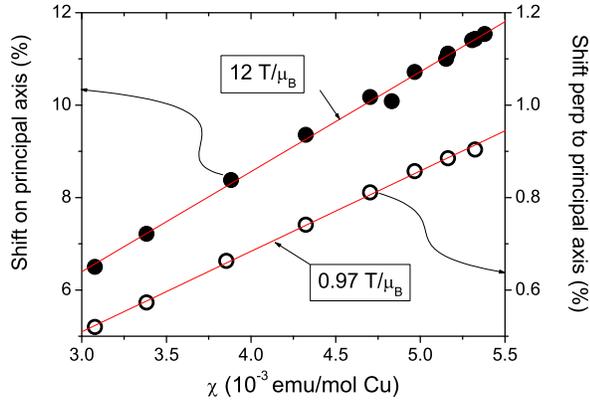}
\caption{\label{fig:BrShift} Br hyperfine shift vs. magnetic
susceptibility with $T$ as an implicit parameter. NMR shifts were
measured on $\nuc{Br}{79}$ at 9\,T between 12 and 50\,K. Black dots correspond to
$K_{zz}$, parallel to the Cu-Br bond, and open circles to
$K_{xx}=K_{yy}$.}
\end{figure}

Having determined the bromine NQR frequencies by experiment, the
remaining unknown parameters in the Hamiltonian given in
Eq.\,\ref{Simplified Hamiltonian} are the Knight shift tensor, the
orientation of the EFG tensor principal axes with respect to the
crystal axes and the associated asymmetry parameter $\eta$.
Several frequency and field scans were performed in the range 10
to 220\,MHz and 5 to 15\,T, respectively, with field applied along
4 different directions, namely [100], [110], [210], and [001]. As
an example, a frequency scan performed with $B_0$=14\,T applied
along [110] is shown in Fig.\,\ref{fig:BrSpectrum}. The computed
NMR frequencies and associated intensities calculated for various
$\eta$ values with $\tens{K}=\tens{0}$ were compared to the
measurements. We concluded that the $Z$-axes of the local EFG
tensors are along the Cu-Br bonds, one of which being oriented
along a direction close to [110] (its exact direction is [1 0.8384
-0.0124]). In addition, a largely anisotropic Knight shift tensor
with its $Z$-axis also parallel to the Cu-Br bonds needed to be
introduced in the Hamiltonian in order to match the computed
frequencies with the measured ones. Furthermore, up to the
precision of our measurements, we deduced that  $\tens{K}$ is
isotropic in the $X-Y$ plane perpendicular to the Cu-Br bond. To
simplify the Hamiltonian, we defined the $X$- and $Y$-axes to be
parallel to the $X$- and $Y$-axes of the EFG tensor. It was then
possible to perform experiments to fully determine the Knight
shift tensor. Indeed, for $\vec{B_0}$ applied along a Cu-Br bond,
that is [1 0.8384 -0.0124], the temperature dependence of the Br
resonance frequencies leads to the determination of $K_{ZZ}$.
Similarly, $K_{XX}=K_{YY}$ can be determined by applying
$\vec{B_0}$ perpendicular to a Cu-Br bond and measuring the
temperature dependence of the Br resonances. By comparing these
measurements to the temperature dependence of the macroscopic
susceptibility, we obtained $K_{ZZ}=12$\,T/$\mu_B$ and
$K_{XX}=K_{YY}=0.97$\,T/$\mu_B$ (see Fig.\,\ref{fig:BrShift}).
From $K_{ZZ}=8/5f_p\mu_B<r^{-3}>$, with
$<r^{-3}>$=$103\cdot10^{24}$cm$^{-3}$\cite{Morton01}, the
experimental $K_{ZZ}$ value leads to $f_p=$4.8\% along the
$Z$-axis. Such a rather large value of $f_{p}$ indicates that the
bromine ligands are involved in the exchange path between Cu
spins.

\begin{figure}
\includegraphics [angle=0, width=3.4in] {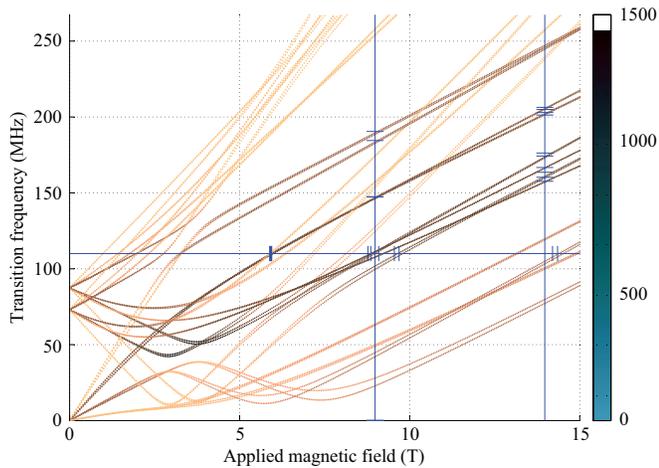}
\caption{\label{fig:BrSim} Computed Br transition frequencies and
intensities as a function of the external magnetic field amplitude
$|\vec{B}_0|$ for $\vec{B}_0$ oriented along [110]. The 48
possible transitions for the two Br isotopes and the four
inequivalent sites are drawn. The intensities are represented on a
color scale in arbitrary units: the lighter the color, the weaker
the intensity. The experimental points corresponding to observed
transitions are denoted by horizontal tips for spectra recorded at
constant field and variable frequencies and by vertical tips for
spectra recorded at fixed frequency sweeping the magnetic field.}
\end{figure}

To determine the only remaining unknown parameter $\eta$, we used
a modified version of the MATLAB routine designed  to minimize the
difference between the measured resonance frequencies and fields,
and the fitted frequencies and fields with $\eta$ as free
parameter. This led to $\eta=0.25\pm0.01$. It should be noted that
we had to take into account a slight misalignment of the crystal
in the coil since a tilt of just one degree away from a specific
direction results in dramatic frequency shifts. In
Fig.\,\ref{fig:BrSim}, we plot the results of simulations for
$B_0$ aligned along a direction close to [110] (the precise
direction is [1 1 -0.08] and corresponds to an experimental
crystal orientation, which was estimated from the comparison
between the measurements shown in Fig.\,\ref{fig:BrSpectrum} and
the calculations). The field dependence of the 48 transitions of
the 2 bromine isotopes located on the 4 inequivalent sites are
plotted. The calculated intensities are shown on a color scale
shown on the right of the figure. On top of the calculated
transitions, horizontal and vertical bars are sketched at the
frequencies, respectively fields, of the observed resonances
measured at fixed field (9\,T and 14\,T), respectively fixed
frequency (110\,MHz). The bars reported at 14\,T correspond to the
measurements shown in Fig.\,\ref{fig:BrSpectrum}.

Although it was not the purpose of the present work to detect all
the transitions, many of them had to be measured in order to
correctly interpret the data and to accurately determine the
unknown parameters in the Hamiltonian. The observed line intensity
ratios do not exactly match the calculated intensity ratios.The
reason for this discrepancy is related to the large variations and
short spin-spin relaxation times (typically on the order of
5-15\,$\mu$s at 15\,K). It should also be noted that the field
dependence of the NMR frequencies of a spin 3/2 with large
quadrupolar couplings placed in a strong external field has
already been numerically calculated using the Liouvillian
formalism and the results were compared to measurements performed
in a $\nuc{Cl}{35}$-sodium chlorate NMR study\cite{Khasawneh02}.
However, the intensity ratios of the transitions were not computed
in this previous study.


\section{Conclusions}

The temperature dependence of the Te NMR relaxation rate clearly
demonstrates the three-dimensional nature of the magnetic phase
transition. This implies that intertetrahedral interactions along
the $c$-axis as well as those in the $ab$-plane are important. The
transition temperature was found to be 10.5\,K at 9\,T. A Br NMR
and NQR study in the paramagnetic phase of Cu$_2$Te$_2$O$_5$Br$_2$
allowed us to demonstrate the important role of bromine in the
interaction paths between Cu spins. In addition,  via tellurium
NMR,  we showed that the [TeO$_3$E] tetrahedra participates in
binding the Br atoms.

The theoretical modelization of this frustrated spin system, topic
of several recent publications, will be clearly facilitated by
this new information. A complete NMR study of this material in its
magnetic phase is currently in progress and is expected to shed
light on its complex magnetic phase.


\begin{acknowledgments}

This work was supported by the  French ANR Grant No. 06-BLAN- 0111.
We thank Fr\'{e}d\'{e}ric Mila and Valeri N. Kotov for helpful
discussions.

\end{acknowledgments}

\bibliography{GlobalBiblio}

\end{document}